# NeuraLunaDTNet: Feedforward Neural Network-Based Routing Protocol for Delay-Tolerant Lunar Communication Networks

*Parth Patel, Milena Radenkovic*
*{psxpp5, milena.radenkovic} @ nottingham.ac.uk*

*School of Computer Science, The University of Nottingham, Nottingham NG8 1BB, UK*

**Abstract**

Space Communication poses challenges such as severe delays, hard-to-predict routes and communication disruptions. The Delay Tolerant Network architecture, having been specifically designed keeping such scenarios in mind, is suitable to address some challenges. The traditional DTN routing protocols fall short of delivering optimal performance, due to the inherent complexities of space communication. Researchers have aimed at using recent advancements in AI to mitigate some routing challenges [9]. We propose utilising a feedforward neural network to develop a novel protocol NeuraLunaDTNet, which enhances the efficiency of the PRoPHET routing protocol for lunar communication, by learning contact plans in dynamically changing spatio-temporal graph.

## 1. Introduction

The exploration of space and the establishment of communication networks beyond Earth's atmosphere presents both immense opportunities and formidable challenges. This research delves into lunar communication, which encompasses rovers and satellites in orbit around the Moon, as well as a ground station on Earth. We explore the integration of a Feedforward Neural Network, developed using PyTorch, which we trained using data generated from the report of simulation using PRoPHET router. This approach aims to harness the power of artificial intelligence to optimize routing decisions. Using a feedforward neural network with only a few layers, is more efficient than using more advanced Deep Learning based approaches, due to the computational and time constraints in lunar communication. However, Graph Neural Networks could be an interesting venue for exploration in future, as its architecture is very appropriate for modelling networks, without using much more computational power.

Our analysis is performed using the Opportunistic Network Environment (ONE) simulator, facilitating a detailed emulation of the conditions and constraints characteristic of Delay-Tolerant Networking (DTN) communication scenarios. A pivotal element of our analysis involves incorporating external mobility traces to authentically model the movements of orbiters.

Deep Java Library has been used to integrate the PyTorch model (made in Python) into the ONE simulator's ProphetRouter.java.

**Limitations of TCP/IP stack**

TCP/IP stack is suitable for the traditional terrestrial scenarios where the network topology is relatively stable, but it faces some issues when the topology of network is dynamic and unstable, despite of improvements like Mobile IP. This is due to the fundamental assumptions of TCP/IP stack, which are, relatively stable network topology, existence of end-to-end routes, and ability to perform a three way handshake. In 1988, with the release of 4.3BSD-Tahoe Unix OS, congestion control mechanism based on the research of Van Jacobson et al. [0] was integrated into TCP. Dynamic topologies can introduce higher rates of packet loss due to frequent node movement, triggering TCP's congestion control mechanisms, causing bad performance.

**VANET and DTN**

Unlike traditional TCP/IP internet, the protocols of Vehicular Ad-hoc Networks are designed specifically for dynamic (ad-hoc) scenarios which lack well established infrastructure. In VANET, the routing process is decentralized, as every node in the network acts as a router.

But they are not very scalable, because maintaining an up-to-date routing table and managing the dynamic topology becomes very complex and resource intensive as the network becomes larger. Moreover, if a node moves out of range or is no longer available, the established path breaks, leading to a disruption in communication. This necessitates the immediate recalibration of routes.

In 2003, the research of Vint Cerf and Kevin Fall led to the development of Delay Tolerant Network architecture [1] for communication scenarios which face severe delays and disruptions. It takes inspiration from ad-hoc network's decentralised architecture but takes it to the next level by adding the store and forward functionality. In the DTN, the traditional internet stack is extended by adding a Bundle Layer. The Bundle Layer implements Bundle Protocol which is responsible for handling store and forward, custody transfer, bundle fragmentation/reassembly, hop-to-hop acknowledgements, etc.

## 2. Literature Review

**General Classification of Strategies: Forwarding vs Replicating**

In general the routing protocols can be classified into two categories [2] : Replication based and Forwarding based. This is more of a spectrum rather than categories.

In replication, multiple copies of same packet are disseminated across multiple paths simultaneously. This increases the probability of delivering message to the destination, but creates overhead and leads to increased use of resources (such as nodes' buffers). Epidemic is the most extreme example of this approach. Message is transferred to every node which comes in contact with a node which contains the message. This is comparable to how diseases spread.

In forwarding, only one copy of packet is transferred from one node to another, and the source node deletes the packet after transfer. This means, at a time, only one node holds a given packet. Examples of this approach are Direct Delivery, First Contact, as well as single-path algorithms like modified Dijsktra. Efficient Singlepath routing is feasible only when it is possible to predict network topology and contact schedule in advance.

There are many algorithms which combine both approaches. For example, Jones et al. [3] proposed selecting path with minimum expected estimated delay, but the data about network topology gets disseminated using epidemic.

**Benchmark Protocols**

*Contact Graph Routing*

In the Space Communication, the contacts are scheduled. That is, we have some apriori knowledge of the schedules of contacts between nodes. Burleigh et al.'s Contact Graph Routing [4], which works by efficiently processing the contact plan, has become the de-facto routing protocol for Space Networks [5].

*PRoPHET (Probabilistic Routing using History of Encounters and Transitivity)*

Each nodes maintains a summary vector which contains delivery predictability to each node. Delivery predictability is calculated based on encounter history.

When node A encounters node B, the delivery predictability to node B in node A's summary vector is updated as follows (and the same is done for delivery predictability to node A in node B's summary vector) :

$P(A,B) = P(A,B) + (1-P(A,B)) \times P_{init}$

Where $P_{init}$ is the initial delivery predictability assigned during first contact.

Transitive updates are also made. This means, if node A meets node B, and node B has higher delivery predictability to node C than node A does, then the delivery predictabily for C in A's summary vector is updated as follows:

$P(A,C) = P(A,C) + (1-P(A,C)) \times P(A,B) \times P(B,C) \times \beta$

Ageing of delivery predictability values is also performed. The rationale behind this is that, if 2 nodes haven't met for a while, it would become increasingly more likely that the network topology has changed and hence it's less likely that they would come in contact.

$$P(A,B) = P(A,B) \times \gamma^{\Delta t}$$

Where gamma is ageing constant, and delta t is the time since last encounter.

Now, when the routing decision is to be made, when a node comes in contact with another node and that other node has higher delivery predictability to a destination than the current node does, then the message gets forwarded from current node to the other node.

## RAPID

Resource Allocation Protocol for Intentional DTN by Balasubramanian et al. [6] assigns a utility value to each message. Utility function is designed to calculate utility value based on factors like ttl, delivery probability, hop count, etc. Messages with higher utility are allocated more resources (like buffer space and bandwidth) than those with lower utility. The goal is to make the most out of limited resources by allocating them to messages which have higher utility. When a node encounters another node, they share the summary vectors which contain the messages present in respective node (and also the utility values of those messages). Based on the utility value, it is decided whether the message should be forwarded to the other node.

## Other Approaches

Making DTN routing more efficient has been an ongoing interest of researchers in this field.

Long K, et al [7] presented the use of a modified Dijkstra algorithm for routing. Network is represented as a graph with nodes as the vertices. Based on the contact information of nodes, the weights of edges of graph are calculated. So essentially, the weights indicate time remaining until next meet. Applying Dijkstra on the graph would give the path with smallest delivery time. This approach requires an apriori knowledge of the network topology. Metrics other than remaining time until contact can also be used. Jain et al. Discussed various metrics which could be used for minimization.

Some works have applied AI techniques for DTN routing. Hylton et al. [8] showed how reinforcement learning can be used. Each node in network maintains a Q table. The rows (states) of Q table are neighboring nodes of current node. Columns (actions) are all the destination nodes in network. Q values are based on estimated delivery time of when routing is done through that particular node to a given destination node. To make routing decision, the state with minimum Q value corresponding to the given destination, is picked. The Q value is updated based on the response received from the picked node. Response would contain better estimate of delivery time.

This way the Q values would continuously improve, as the nodes closer to destination would return better and better estimates of delivery time.

Dudukovich et al. [9] discussed limitations of RL based approach. Defining an appropriate reward function is not very feasible in DTN scenario. The reward function could be based on either estimated delivery time (as described above) and/or on estimated delivery probability. Relying on delivery time could be unreliable as the delays and disruptions could often be impossible to predict. Relying on delivery probability is unfeasible during a continual learning approach, as this would require end-to-end delivery acknowledgement, which is hard to implement in DTN.

Dudukovich et al. [9] elaborates a way to train traditional ML models like DecisionTree to learn successful routing strategies. The approach treats routing as a multi-class classification problem. The authors trained their model on ZebraNet mobility traces. The following are input features used: Time index in the epoch, Source node, Destination node, Region code for source node, Region code for destination node, and Message delivery status. The classifier returns the output labels corresponding to each node indicating if the message was forwarded to that node.

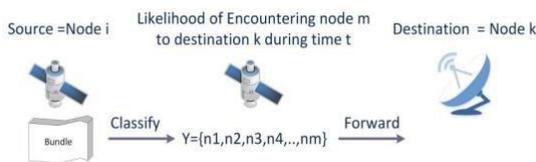

*(Figure 1: DTN Routing as classification problem [9])*

There have recently been proposals [10] to explore the potential of using Neural Processes (which are a combination of neural net and probabilistic modelling) for routing. Neural Processes (NPs) are already used in climate modeling, but their applications remain unexplored in the field of routing research. It could open up possibilities for tremendous new advancements. NPs can be modelled to predict future disruptions based on historical data. The predictions can be used to make real time routing strategies. One of the key strengths of NPs is their ability to handle variable and complex patterns of missing data, a common issue in DTNs due to their unpredictable connectivity. This capability allows NPs to maintain a high level of performance even when data inputs are incomplete or sporadic, which is essential for networks operating in remote or unstable environments. DTNs often deal with irregularly sampled spatio-temporal data, and NPs excel in this area. They can effectively process and interpret data that is non-uniformly distributed in time and space, providing reliable insights and predictions that are crucial for the efficient functioning of DTNs. This ability is particularly valuable in managing network resources and routing in unpredictable environments.

## 3. Methodology for generating training data for NeuraLunaDTNet

The lunar communication setup consists of 3 groups: Moon rovers, Moon orbiters, and Earth ground station. Rovers and orbiters on the moon try to send messages (which would in real life contain some data pertaining to scientific studies) to earth ground station. As the Earth ground station could be out of sight for rovers and orbiters on the far-side of Moon, the messages would get relayed to orbiters and rovers on the near-side.

For Earth ground station, StaticMobility Model of The ONE simulator was used. For the former two, ExternalMobility Model was used with the traces mentioned below (in the *Satellite Mobility Traces* sub-section). In total there are 151 nodes. The routers used were PRoPHET and Epidemic, and two different buffer sizes were tried for both. Each simulation runs for 30 minutes.

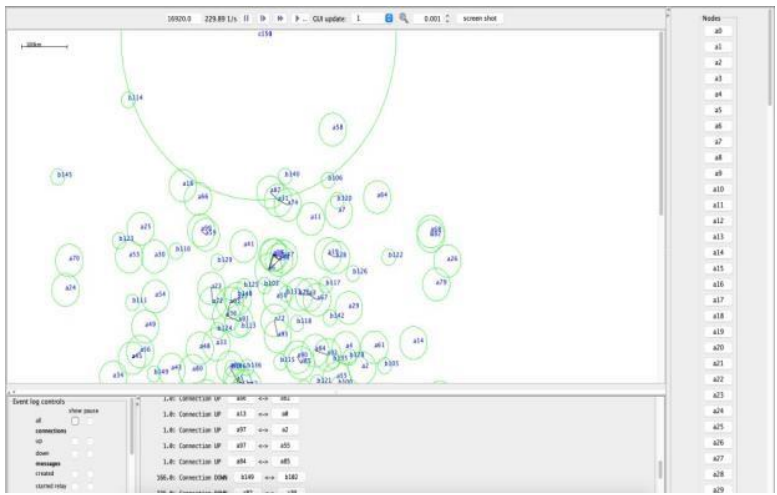

*(Figure 2: Graphical output of ONE Simulation)*

### Satellite Mobility Traces

A dataset of mobility patterns of 600 satellites collected over 7 days has been used. This dataset was published as part of the International Data Analytics Olympiad 2020 (IDAO 2020) Competition. To convert the data to ONEcompatible mobility traces, we developed a conversion tool in Python. Since the original dataset is 3D, projection had to be performed. Binning of the time variable was also performed, so rather than being a continuous value, it would be an epoch between 0 and 167 (so basically an epoch indicates 1 hour in each of the 7 days). The size of world space was mapped down from 15.44 million KM^2 to about 15,44,400 KM^2.

### Generating Training Data

Our core aim was to train a neural network based on a report generated by simulation, in order to learn routing. For this, we coded a new type of report – NNTrainerReport.java, by modifying the DeliveredMessagesReport.java. Usually each tuple's 1$^{st}$ cell contains the time at which message was delivered. We changed it to contain time at which message was created.

```java
public void messageTransferred(Message m, DTNHost from, DTNHost to,
        boolean firstDelivery) {
    if (!isWarmupID(m.getId()) && firstDelivery) {
        int ttl = m.getTtl();
        write(format(m.getCreationTime()) + " " + m.getId() + " " +
                m.getSize() + " " + m.getHopCount() + " Y " +
                format(getSimTime() - m.getCreationTime()) + " " +
                m.getFrom() + " " + m.getTo() + " " +
                (ttl != Integer.MAX_VALUE ? ttl : "n/a") +
                (m.isResponse() ? " Y " : " N ") + getPathString(m));
    }
}
```

*(Figure 3: Code of NNTrainerReport.java)*

Other than the NNTrainerReport, MessageStatsReport was also generated.

Here's a comparison between various reported metrics for PRoPHET and Epidemic simulations, both for two different buffer sizes. The simulation was run for 30 minutes (for each of the four).

| Epidemic | 50M Buffer | 100M Buffer |
|---|---|---|
| Messages Created | 244 | 244 |
| Messages Started | 28010009 | 25157384 |
| Messages Relayed | 28009984 | 25157351 |
| Messages Dropped | 28000320 | 25137815 |
| Messages Delivered | 67 | 131 |

| PRoPHET | 50M Buffer | 100M Buffer |
|---|---|---|
| Messages Created | 244 | 244 |
| Messages Started | 50957132 | 35157948 |
| Messages Relayed | 50957088 | 35157901 |
| Messages Dropped | 50947283 | 35138328 |
| Messages Delivered | 93 | 153 |

PRoPHET started, relayed and dropped less messages than Epidemic, as expected.

For the sake of neural network training, only the report of PRoPHET with large buffer size would be used.

## 4. Training the Neural Network based Protocol NeuraLunaDTNet

Pandas library was used to convert the report to an appropriate format for training data. The values in path column (which contains names of hop nodes separated by '->') were split and used to create tuples which contain current node and next hop.

The custom neural network architecture contains 4 neurons on input layer, for: SourceNode, DestinationNode, Epoch, CurrentNode.

The output layer has just 1 neuron which would contain the regressed node ID of the next hop to which the message shall be forwarded. Since this is to be treated as a regression task, no activation function has been used on output layer and last hidden layer.

PyTorch library has been used for creating the model and training. The below image shows the architecture of the neural network. It contains 4 hidden layers.

```python
class ACNet(nn.Module):
    def __init__(self):
        super(ACNet, self).__init__()
        self.fc1 = nn.Linear(4, 64)
        self.fc2 = nn.Linear(64, 512)
        self.fc3 = nn.Linear(512, 128)
        self.fc4 = nn.Linear(4096, 256)
        self.fc5 = nn.Linear(128, 6)
        self.fc6 = nn.Linear(6, 1)

    def forward(self, x):
        x = F.relu(self.fc1(x))
        x = F.relu(self.fc2(x))
        x = F.relu(self.fc3(x))
        #x = F.relu(self.fc4(x))
        x = F.relu(self.fc5(x))
        #x = F.relu(self.fc6(x))
        x = self.fc6(x)
        return x

# NN, Loss Function, and Optimizer
net = ACNet()
criterion = nn.MSELoss()
optimizer = optim.Adam(net.parameters(), lr=0.007)
```

*(Figure 4: Code showing the architecture of neural network)*

The training runs for around 70k epochs, and during each epoch, the mean squared error of regressed output of neural network is calculated w.r.t. the real output in training data. Partial derivatives of mean squared error are calculated w.r.t. each weight and bias in each of the neurons in network, and subtracted from respective weights and biases (this process is called gradient descent).

Here's how the Mean Squared Error dropped during the training:

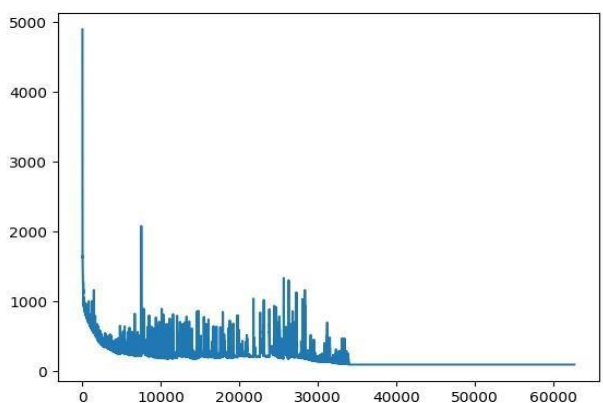

*(Figure 5: Drop of MSE over training)*

It stagnated at around 100. From inspecting the outputs, we noticed that the regressed node ID's were very accurate, but with errors of +/- 1 (which means if the neural network regressed the node ID for next hop as x, it could in reality be x-1 or x+1).

The trained model was saved in a .pt file.

**Integrating PyTorch model into ONE simulator**

Deep Java Library has to be used in order to be able to load a saved PyTorch model in Java. Firstly, the ONE simulator's Java project was converted into Maven project, so that additional dependencies can be managed effectively. Following special dependencies were added to the pom.xml file:

- PyTorch Engine (ai.djl.pytorch:pytorchengine:0.25.0)
- PyTorch Native CPU (ai.djl.pytorch:pytorchnative-cpu:2.1.1)
- DJL API (ai.djl:api:0.25.0)

Once this is done, the required namespaces of DJL package are imported in the ProphetRouter.java file. The tryOtherMessages method is modified by adding following code:

```java
if (othRouter.hasMessage(m.getId())) {
    continue; // skip messages that the other one has
}
if (othRouter.getPredFor(m.getTo()) > getPredFor(m.getTo())) {
    this.model = Model.newInstance("PyTorch");

    try {
        this.model.load(Paths.get("/Users/zetro7744/Downloads/savedmodel2.pt"));
    } catch (MalformedModelException | IOException e) {
        // TODO Auto-generated catch block
        e.printStackTrace();
    }

    this.manager = NDManager.newBaseManager();
    this.translator = new SimpleTranslator();
    this.predictor = this.model.newPredictor(this.translator);

    List<DTNHost> hops = m.getHops();
    float nextHopID = 0; //need to implement logic to get ID of otherRouter
    float epoch = (float) Math.floor(m.getCreationTime()/60000);
    float deliveringFrom = Float.parseFloat(m.getFrom().toString().substring(1));
    float deliveringTo = Float.parseFloat(m.getTo().toString().substring(1));
    float currentNode = Float.parseFloat(hops.get(m.getHopCount() - 1).toString().substring(1));
    float[] floatData = new float[]{epoch, deliveringFrom, deliveringTo, currentNode};
    FloatBuffer floatBuffer = FloatBuffer.wrap(floatData);
    NDArray inputArray = this.manager.create(floatBuffer, new Shape(1, floatData.length), DataType.FLOAT3
    NDList result;
    try {
        result = this.predictor.predict(new NDList(inputArray));
        NDArray outputArray = result.singletonOrThrow();
        float output = outputArray.getFloat();
        if(nextHopID > output - 5 && nextHopID < output - 5) {
            messages.add(new Tuple<Message, Connection>(m,con));
        }
    }
    catch (TranslateException e) {
        e.printStackTrace();
    }
}
```

*(Figure 6)*

It loads the model, feeds into it the inputs (epoch, source node, dest node, current node), and if the model gives output which is +/-5 of the neighbouring node to which the PRoPHET is considering forwarding message to, then the message gets forwarded.

# 5. Conclusion and Future Work

In this study, we demonstrated the potential of leveraging Deep Learning, specifically through a PyTorch-trained feedforward neural network, to enhance the PRoPHET protocol for interplanetary communication networks. This integration not only showcases the feasibility but also the benefits of incorporating Deep Learning techniques in the realm of space communications. Our findings suggest a promising avenue for making interplanetary communication more efficient, which is critical for the success of future space exploration missions.

While the feedforward neural network implemented by us relied only of some basic input features like message creation epoch, source node, destination node, etc., there's a scope to make the neural network more sophisticated by making it capable of taking into consideration potential disruptions, erretic path changes due to certain events, and so on. Moreover, more advanced Neural Network architectures, such as Graph Neural Networks (which could be very appropriate for space network modelling) could be explored. We aim to incorporate our novel protocol in the MODiToNeS platform [11], and deploy it in the next generation space ad-hoc networks [12].